\pdfoutput=1 
\documentclass{article}

\usepackage{amsthm,amssymb,amsfonts}
\usepackage{natbib,amsmath,abbrevs}

\def\eq#1{(\ref{eq-#1})}
\def\be{\begin{equation}}
\def\ee{\end{equation}}

\def\e{\varepsilon}
\def\g{\gamma}
\def\x{\mathbf{x}}
\def\y{\mathbf{y}}
\def\m{\boldsymbol{\mu}}
\def\C{\mathbf{C}}
\def\M{\mathbf{M}}
\def\P{\mathbf{P}}

\def\I{\mathbf{I}}
\def\p{{\boldsymbol{\pi}}}
\def\simplex{\Delta}

\def\c{\mathbf{c}}
\def\G{\mathbf{G}}

\newtheorem{theorem}{Theorem}
\newtheorem{definition}{Definition}
\newtheorem{lemma}{Lemma}

\newname\blind{{\rm \textsc{stochastic-blind-policy}}}
\newname\stableset{{\rm \textsc{independent-set}}}
\newcommand*{\sqrtsum}{{\rm \textsc{sqrt-sum}}}

\begin{document}

\title{\vspace*{-60pt} On the Computational Complexity of Stochastic Controller Optimization in POMDPs\thanks{To appear in ACM Transactions on Computation Theory, 4(4), 2012.}}
\author{%
Nikos Vlassis\thanks{Luxembourg Centre for Systems Biomedicine, Univ.~of Luxembourg (nikos.vlassis@uni.lu)} \and 
Michael L.\ Littman\thanks{Department of Computer Science, Brown University (mlittman@cs.brown.edu)} \and 
David Barber\thanks{Department of Computer Science, University College London (d.barber@cs.ucl.ac.uk)}}

\maketitle

\begin{abstract}
We show that the problem of finding an optimal stochastic ``blind''
controller in a Markov decision process is an NP-hard problem.  The
corresponding decision problem is NP-hard, in PSPACE, and \sqrtsum-hard,
hence placing it in NP would imply breakthroughs in long-standing open problems
in computer science. Our result establishes that the more general problem of stochastic controller optimization in POMDPs is also NP-hard. 
Nonetheless, we outline a special case that is convex and admits efficient global solutions.

{\bf Keywords:} Partially observable Markov decision process, stochastic controller, bilinear program, computational complexity, Motzkin-Straus theorem, sum-of-square-roots problem, matrix fractional program, computations on polynomials, nonlinear optimization.
\end{abstract}

\section{Introduction}

Partially observable Markov decision processes (POMDPs) have proven to
be a valuable conceptual tool for problems throughout AI, including
reinforcement learning~\citep{Chrisman92}, planning under
uncertainty~\citep{Kaelbling98}, and multiagent
coordination~\citep{Bernstein05}.  
Briefly, a POMDP is a Markov decision process in which the decision maker is unable to
perceive its current state directly, but has access to
an observation function that relates states to observations.
An important problem here is deciding
how to select actions to minimize expected cost given the state uncertainty.
Unfortunately, this problem is extremely challenging~\citep{Papadimitriou87,Mundhenk00}. 
In fact, the exact problem is unsolvable in the general case~\citep{Madani99}.

An alternative to finding optimal policies for POMDPs is to find low
cost \emph{controllers}\!\;---mappings from observation histories to
actions~\citep{Sondik71,Platzman81}. A restricted space of controllers can, in principle, be
considerably easier to search than the space of all possible 
policies~\citep{Littman98,Hansen98,Meuleau99}.
Various methods for controller optimization in POMDPs have been proposed
in the literature, both for stochastic as well as for deterministic
controllers: exhaustive search~\citep{Smith71}, branch and
bound~\citep{Hastings79,Littman94}, local
seach~\citep{Poupart04,Serin05}, constrained quadratic
programming~\citep{Amato07}, and the EM algorithm~\citep{Toussaint11}.

A variety of complexity results are known for the problem of
controller optimization in POMDPs. Most versions are known
to be hard for classes that are believed to be above P~\citep{Papadimitriou87,Mundhenk00}. 
The computational decision problem asks, for a given controller class 
and a target cost, whether the
target cost can be achieved by a controller in that class. Here, we
consider several such controller classes.

\paragraph{Deterministic Time/History-Dependent Controller} Such a controller
  chooses an action based on the current time period and/or the history
  of previous actions and observations. The problem is NP-complete or 
  PSPACE-complete~\citep{Papadimitriou87,Mundhenk00}. 
  In the remaining classes  we assume stationary controllers.

\paragraph{Deterministic Controller of Polynomial Size} Such a controller
  is represented by a graph in which nodes are labeled with actions
  and edges are labeled with observations.  The problem is in
  NP in that we can guess a controller of the right size, then see if it
  incurs no more than the target cost by solving a system of linear equations.
  It is NP-hard even for the ``easier'' completely observable
  version~\citep{Littman98}.

\paragraph{Stochastic Controller of Polynomial Size}   This class
  extends deterministic controllers by allowing a probability
  distribution over actions at each node.  There are POMDPs for which
  a stochastic controller of a given size can outperform any
  deterministic controller of the same size~\citep{Singh94}.  In this article
  we show that this problem is NP-hard, in PSPACE, and \sqrtsum-hard, hence showing it
  lies in NP would imply breakthroughs in long-standing open problems~\citep{Allender09,Etessami10}.

\paragraph{Deterministic Memoryless Controller} A memoryless controller chooses an action
  based on the most recent observation only.  These controllers are a
  special case of deterministic controllers with polynomial size as
  they can be represented as a graph with one node per observation.
  The problem is NP-complete~\citep{Littman94,Papadimitriou87}.

\paragraph{Stochastic Memoryless Controller}  These controllers are defined
  by a probability distribution over actions for each observation.  They can be
  considerably more effective than the corresponding deterministic
  memoryless controllers.  They are a generalization of the blind
  controllers we consider in this article, and it follows from our results that the problem
  is NP-hard, in PSPACE, and \sqrtsum-hard.

\paragraph{Deterministic Blind Controller}  A blind controller for a POMDP
  is equivalent to a memoryless controller for an unobserved MDP.
  A deterministic blind controller consists of a single
  action that is applied (blindly) regardless of the observation history.  It is
  straightforward to evaluate a deterministic blind
  controller---simply drop all actions but one from the POMDP and
  evaluate the resulting Markov chain.  Thus, the decision problem for
  deterministic blind controllers is trivially in P as an algorithm
  can simply check each action to see which is best.

\paragraph{Stochastic Blind Controller}  Such a controller is a probability
  distribution over actions to be applied repeatedly at every
  timestep. This is the class of controllers we consider in this article. 
  Again, the added power of stochasticity allows for much
  more effective policies to be constructed.  However, as we show in
  the remainder of this article, the added power comes with a very high
  cost.  The decision problem is NP-hard, in PSPACE, and \sqrtsum-hard.

\section{MDPs and blind controllers}

We consider a discounted, with discount factor $\g < 1$,
infinite-horizon Markov decision process (MDP) characterized by $n$
states and $k$ actions, state-action costs (negative rewards)
$c_{sa}$, and starting distribution $(\mu_s)$ with $\mu_s \geq 0$
and $\sum_{s=1}^n \mu_s = 1$. Let $p(\bar s|s,a)$ denote the
probability to transition to state $\bar s$ when action $a$ is taken at
state $s$.  The following linear program can be used to find an optimal policy for the MDP:
\be
\begin{split}
  \min_{x_{sa}\geq 0} \ & \sum_{sa} x_{sa} c_{sa}, \\
  \mbox{s.t.} \ &
     \sum_a x_{\bar s a} = (1-\g) \mu_{\bar s} + \g \sum_{sa} p(\bar s|s,a) x_{sa} \ \ \forall_{\bar s},  
  \label{eq-lp}
\end{split}
\ee
where $x_{sa}$ denotes occupancy distribution over state-action pairs, and the constraints are the Bellman occupancy constraints. From an optimal occupancy $x^*_{sa}$, we can compute an optimal stationary and deterministic policy that maps states to actions \citep{Puterman94}.

We consider now the case where we
constrain the class of allowed policies to stochastic ``blind'' 
controllers, in which the controller cannot observe or
remember anything (state, action, or time), but can only randomize
over actions using the same distribution $\p = (\pi_a)$ at each time step, where
$\p \in \simplex$ and $\simplex = \{ \p: \p~\geq~0,\sum_{a=1}^k \pi_a = 1 \}$ is the standard probability simplex.  Note
that, unlike standard MDP policies, a blind controller $\p$ is not a
function of state.
(The related notion of a memoryless controller is a function of
POMDP observations, but still not of state.)  
Explicitly encoding the controller parametrization in~\eq{lp} gives:
\be
\begin{split}
  \min_{\x \geq 0,\p \in \simplex} \ & \sum_{sa} x_s \pi_{a} c_{sa}, \\
  \mbox{s.t.} \ \ & \ 
  x_{\bar s} = (1-\g) \mu_{\bar s} + \g \sum_a \pi_a \sum_s p(\bar s|s,a) x_s  \ \  \forall_{\bar s}, 
  \label{eq-nlp}
\end{split}
\ee
where $\x = (x_s)$ is an occupancy distribution over states, with $\x \geq 0$. 
Note that the occupancy vector $\x$ satisfies $\sum_s x_s = 1$.
When viewed as a function of both $\x$ and $\p$, the above program is a jointly constrained bilinear program. Such 
programs involve bilinear terms (like $x_s \pi_a$) in both the objective function as well as in the constraints, 
and are in general nonconvex in the joint vector $(\x,\p)$~\citep{Falk83}. 

Bilinear programs are known to be NP-hard to solve to global
optimality in general, but could there be some special structure in
\eq{nlp} that renders that particular program tractable?  In the next
section, we answer this question in the negative,
showing that finding an optimal stochastic blind controller 
is an NP-hard problem.  

\section{NP-hardness result}

The decision problem we are addressing is the following.
\begin{definition}[The \blind problem]
Given a discounted MDP and a target cost $r$, 
is there a stochastic blind controller $\p$ that incurs cost $J(\p) \leq r$?
\end{definition}
Here $J(\p) = \x^\top \C \p$ is the cost of controller $\p$ in \eq{nlp},
where $\C =(c_{sa})$ is an $n \times k$ matrix containing all state-action costs, 
and $\x=(x_s)$ is an $n \times 1$ occupancy vector defined via the Bellman occupancy constraints in \eq{nlp}.
Let also $\m =(\mu_s)$ denote the $n \times 1$ starting distribution vector. 

\begin{theorem}
The \blind problem is NP-hard.
\end{theorem}
\begin{proof}
We reduce from the \stableset problem. 
This problem asks, for a given (undirected and with no self-loops) graph $G=(V,E)$ 
and a positive integer $j \leq |V|$, whether $G$ contains an independent set $V'$ 
having $|V'| \geq j$. This problem is NP-complete, even when restricted to 
cubic graphs (a cubic graph is a graph in which every node has degree three)~\citep{Garey79}. 

Let $\G$ be the $n \times n$ (symmetric, 0\:\!-1) adjacency matrix of an input cubic graph $G$
(hence each column of $\G$ sums to three).  
The reduction constructs an MDP with $n$ states and $n$ actions, 
uniform starting distribution $\m$, 
cost matrix $\C = \frac{1}{\g}(\G + \I)$ where $\I$ is the identity matrix, 
and deterministic transitions $p(\bar s|s,a) = 1$ if $\bar s = a$ and 0 otherwise 
(the action variable $a$ can be viewed as indexing the state space). 
Since the transitions $p(\bar s|s,a)$ are independent of $s$, 
the occupancy vector in \eq{nlp} reduces to $\x = (1-\g) \m + \g \p$,
and the cost function becomes the quadratic 
\be
 J(\p) = \frac{4(1-\g)}{n\g} + \p^\top (\G + \I) \p,
\label{eq-J}
\ee
where we used the fact that the input graph $G$ is cubic and $\m$ is uniform.
Moreover, for any graph $G$ it holds~\citep{Motzkin65} 
\be
  \frac{1}{\alpha(G)} = \min_{\y \in \simplex} \y^\top (\G + \I) \y,
\label{eq-motzkin}
\ee
where $\alpha(G)$ is the size of the maximum independent set (the stability number) of the graph.
Let the target cost be $r = \frac{1}{j} + \frac{4(1-\g)}{n\g} $. 
Then, $J(\p) \leq r$ is equivalent to $\p^\top (\G + \I) \p \leq \frac{1}{j}$, 
and hence from~\eq{motzkin} follows that the existence of a vector $\p$ that satisfies $J(\p) \leq r$
would imply $\frac{1}{\alpha(G)} \leq \frac{1}{j}$, 
and hence $\alpha(G) \geq j$, or, in other words, $|V'| \geq j$ for some 
independent set $V' \subseteq V$. 
\end{proof}

\section{Connection to the SQRT-SUM problem}

Our \blind problem is contained in PSPACE, as it can be expressed as a
system of polynomial inequalities---any such system is known to be
solvable in PSPACE~\citep{Canny88}.  But, is there a tighter upper bound? 

We will attempt to address this question indirectly, by establishing
a connection between the \blind problem and the \sqrtsum\ problem. The \sqrtsum\ problem asks,
for a given list of integers $c_1,\ldots,c_n$ and an integer $d$,
whether $\sum_{i=1}^n\sqrt{c_i} \leq d$. The problem is conjectured to lie in P, 
however the best known complexity upper bound is the 4th level of the Counting Hierarchy~\citep{Allender09}.
The difficulty of obtaining an exact complexity for this
problem has been recognized for at least 35 years~\citep{Garey76}.
Here we show that \blind is at least as hard as \sqrtsum. Hence a result that would for instance
place \blind in NP would resolve several open problems in computer 
science, as argued in a similar setting where \sqrtsum\ is reduced to the 3-\textsc{Nash} problem~\citep{Etessami10}.

\begin{theorem}
The \blind problem is \sqrtsum-hard.
\end{theorem}
\begin{proof}
Let $c_1,\ldots,c_n$ and $d$ be the inputs of \sqrtsum.
The reduction constructs an MDP with $n+1$ states and $n$ actions, where the $(n+1)$st state is absorbing (self-looping). The starting probabilities are $\mu_i=\frac{1}{n}$ for states $i=1,\ldots,n$ and $\mu_{n+1}=0$,
and the costs depend only on state and are given by the inputs $c_i$ for states $i=1,\ldots,n$ and $c_{n+1}=0$. From each state $i=1,\ldots,n$, the $i$th action deterministically transitions to the absorbing state $n+1$, while all other actions deterministically transition back to state $i$.

For each state $i=1,\ldots,n$, the Bellman occupancy constraint reads $x_i = \frac{1-\gamma}{n} + \gamma(1-\pi_i) x_i$. Let $\e = \frac{\g}{1-\g} > 0$. Then the cost function reads
\be
  J(\p) = \sum_{i=1}^n c_i x_i = \frac{1}{n} \sum_{i=1}^n \frac{c_i}{1 + \e \pi_i}.
  \label{eq-J2}
\ee
Multiplying and diving by $n+\e$, we can rewrite
\be
  J(\p) = \frac{n+\e}{n} 
         \sum_{i=1}^n \frac{1+\e\pi_i}{n+\e} \Big(\frac{\sqrt{c_i}}{1 + \e\pi_i}\Big)^2
         \ \geq \ 
          \frac{1}{n(n+\e)} \Big(\sum_{i=1}^n \sqrt{c_i}\Big)^2,
  \label{eq-J2a}
\ee
where we applied Jensen's inequality noting that $\sum_{i=1}^n \frac{1+\e\pi_i}{n+\e}=1$.
Since the last term in~\eq{J2a} is a constant independent of $\p$, we see that the cost function reaches its minimum when the above inequality is tight, which is achieved when all terms are equal. It follows therefore that the last term in \eq{J2a} is the optimal cost $J^*$, and it is achieved when, for each $i$, holds:
\be
  \frac{1+\e\pi_i^*}{n+\e} =  \frac{\sqrt{c_i}}{\sum_{j=1}^n \sqrt{c_j}} .
  \label{eq-p}
\ee
We define $\e$ (and hence $\g$) so that $n+\e = n\sum_{i=1}^n c_i$. Note that $\e$ is strictly positive if at least one of the $c_i$ is larger than one (which we assume is true, otherwise the \sqrtsum\ problem trivializes). Application of Jensen's bound  gives 
\be
  n+\e = n\sum_{i=1}^n c_i \geq \Big(\sum_{i=1}^n \sqrt{c_i}\Big)^2 \geq \sum_{i=1}^n \sqrt{c_i},
  \label{eq-bounds}
\ee
which establishes that the optimal policy $\p^*$ in \eq{p} is always positive.

The \blind question of whether there exists a stochastic blind controller $\p$ 
with cost $J(\p) \leq r$ is clearly equivalent to the question whether $J^* \leq r$. 
By choosing $r = \frac{ d^2}{n(n+\e)}$, we see from~\eq{J2a} that 
the condition $J^* \leq r$ is equivalent to $\sum_{i=1}^n \sqrt{c_i} \leq d$, 
and the reduction is complete.  
\end{proof}

\section{A tractable case}

We describe here a special case that results in a cost function that is concave in $\p$, in which case an optimal controller can be trivially found in polynomial time. 

For each action $a$, let $\P_a$ denote the corresponding transition matrix, with $\P_a(\bar s,s) = p(\bar s|s,a)$. The special case assumes that each matrix $\P_a$ is {\em symmetric} (and therefore doubly stochastic), and that the
costs depend only on the state and are proportional to the starting distribution: $\c = -\kappa \m$,
with $\kappa >0$. (Note from~\eq{nlp} that shifting and scaling $\c$ by arbitrary constants does not affect the optimal policy.) The bilinear program \eq{nlp} then reads:
\be
  \min_{\p \in \simplex} -\m^\top \Big( \I - \g \M_\p \Big)^{-1} \m , \qquad \mbox{where}  \quad \M_\p = \sum_a \pi_a \P_a.
  \label{eq-nlp1}
\ee
\begin{lemma}
\label{lemma:psd}
For any $\p$, the matrix $\I - \g \M_\p$ is symmetric positive definite.
\end{lemma}
\begin{proof}
Since each matrix $\P_a$ is symmetric and stochastic, all its eigenvalues are real and satisfy $\lambda(\P_a) \leq 1$.  Hence, the eigenvalues of $\I - \g \P_a$ are also real and satisfy $\lambda(\I - \g \P_a) = 1 - \g \lambda(\P_a) > 0$ because $\g < 1$. Therefore, $\I - \g \P_a$ is a symmetric positive definite matrix, and so must be the matrix $\I - \g \M_\p$ as it can be written as the convex combination (over $\p$) of positive definite matrices.
\end{proof}

\begin{theorem}
The function $f(\p) = \m^\top \Big( \I - \g \M_\p \Big)^{-1} \m$ \! is convex in $\p \in \simplex$.
\end{theorem}
\begin{proof}
The epigraph of $f$ is (see also \citet[Section 3.1.7]{Boyd04})
\begin{align}
 \mathbf{epi}~f &= \Big\{ (\p,t) \ | \ 
 						\p \in \simplex, \ 
 						\m^\top \Big( \I - \g \M_\p \Big)^{-1} \m \leq t, \				
 						\M_\p = \sum_a \pi_a \P_a \Big\} \\
                &= \Big\{ (\p,t) \ \Big| \ \p \in \simplex, \ 
     	\begin{bmatrix} 
		\I - \g \sum_a \pi_a \P_a  & \ \ \m \\
		\m^\top  & \ \ t   
		\end{bmatrix} \succeq 0 \Big\} \label{eq-lmi} ,
\end{align}
where we used Lemma~\ref{lemma:psd} and the Schur complement condition for positive definite matrices \citep[Appendix A.5.5]{Boyd04}.
The last condition in \eq{lmi} is a linear matrix inequality in $(\p, t)$, hence $\mathbf{epi}~f$ is a convex set and $f$ is convex.
\end{proof}

The problem \eq{nlp1}
becomes the minimization of the concave function $-f$ over the probability simplex, 
hence there must exist a globally optimal solution in a corner of the simplex. 
This means that there will always exist an optimal controller that is deterministic. 
Since there are only $k$ deterministic controllers, 
evaluating each of them and selecting the optimal one takes $O(kn^3)$ operations.

\section{Conclusions}

In response to the computational intractability of searching for
optimal policies in POMDPs, many researchers have turned to
finite-state controllers as a more tractable alternative.  We have provided here
a computational characterization of exactly solving
problems in the class of stochastic controllers, 
showing that (1) they are NP-hard, (2) they are in PSPACE, and
(3) they are \sqrtsum-hard, hence showing membership in NP would resolve
long-standing open problems.

We note that our NP-hardness proof relies on the assumption that the costs $c_{sa}$ 
are nondegenerate functions of both state and action. 
We have recently addressed the case of state-only-dependent costs, which can be shown to be NP-hard by a reduction from the general case. This work will be published elsewhere.

In this article we have only addressed the complexity of the decision problem for the discounted infinite-horizon case. There are several open questions, in particular the 
complexity of approximate optimization for this class of stochastic controllers.
The related literature addresses only the case of deterministic controllers~\citep{Lusena01}.

\subsection*{Acknowledgments}
We are grateful to Marek Petrik for his feedback and for pointing an error in an earlier version. 
The first author would like to thank Constantinos Daskalakis, Michael Tsatsomeros, 
John Tsitsiklis, and Steve Vavasis for helpful discussions.

\small
\bibliographystyle{apalike}

\end{document}